\documentclass[10pt,a5paper,english]{IEEEtran}
\usepackage[T1]{fontenc}
\usepackage[latin9]{inputenc}
\usepackage{amssymb}
\usepackage{graphicx}

\makeatletter

\pdfpageheight\paperheight
\pdfpagewidth\paperwidth

\usepackage{cite}

\makeatother

\usepackage{babel}
\begin{document}

\title{Research on Wireless Multi-hop Networks: Current State and Challenges}

\author{Guoqiang Mao%
\thanks{This research is partially supported by ARC Discovery project DP110100538.
This material is based on research partially sponsored by the Air
Force Research Laboratory, under agreement number FA2386-10-1-4102.
The U.S. Government is authorized to reproduce and distribute reprints
for Governmental purposes notwithstanding any copyright notation thereon.
The views and conclusions contained herein are those of the authors
and should not be interpreted as necessarily representing the official
policies or endorsements, either expressed or implied, of the Air
Force Research Laboratory or the U.S. Government.%
}\\
School of Electrical and Information Engineering\\
The University of Sydney\\
National ICT Australia\\
Email: guoqiang.mao@sydney.edu.au}
\maketitle
\begin{abstract}
Wireless multi-hop networks, in various forms and under various names,
are being increasingly used in military and civilian applications.
Studying connectivity and capacity of these networks is an important
problem. The scaling behavior of connectivity and capacity when the
network becomes sufficiently large is of particular interest. In this
position paper, we briefly overview recent development and discuss
research challenges and opportunities in the area, with a focus on
the network connectivity. 
\end{abstract}
\thispagestyle{empty}

\section{Introduction\label{sec:Introduction}}

Wireless multi-hop networks, in various forms, e.g. wireless sensor
networks, underwater sensor networks, vehicular networks, mesh networks
and UAV (Unmanned Aerial Vehicle) formations, and under various names,
e.g. ad-hoc networks, hybrid networks, delay tolerant networks and
intermittently connected networks, are being increasingly used in
military and civilian applications. There are three defining features
that characterize a wireless multi-hop network:
\begin{enumerate}
\item Wireless devices are self-organized or assisted by some infrastructure
to form a network. The former case corresponds to ad-hoc networks
whereas the latter case corresponds to infrastructure-based multi-hop
networks. Depending on the applications, the forms of the infrastructure
can be quite flexible, e.g. a subset of devices connected via wired
connections, a subset of devices with more powerful transmission capability
such that they form a wireless backbone for the network, or in a UAV
formation, the infrastructure may assume the form of a subset of UAVs
with satellite links.
\item Communication is mostly via wireless multi-hop paths. This feature
sets wireless multi-hop networks apart from the traditional one-hop
networks, i.e. cellular networks and wireless LANs. Therefore, there
is a unique set of challenging problems specific to wireless multi-hop
networks.
\item Packets are forwarded \emph{collaboratively} from the source to the
destination.
\end{enumerate}
Studying connectivity and capacity of wireless multi-hop networks
is an important problem \cite{Gupta98Critical,Haenggi09Stochastic,Penrose03Random}.
The scaling behavior of connectivity and capacity when the network
becomes sufficiently large is of particular interest. In this paper,
we briefly overview recent development and discuss research challenges
and opportunities in the area, with a focus on the network connectivity.
A network is said to be \emph{connected} iff (if and only if) there
is a (multi-hop) path between any pair of nodes. Further, a network
is said to be $k$-connected if there are $k$ mutually independent
paths between any pair of nodes that do not have any node in common
except the starting and the ending nodes. $k$-connectivity is often
required for robust operations of the network.

The rest of the paper is organized as follows: Section \ref{sec:Large scale network}
discusses connectivity of large-scale random networks; Section \ref{sec:Giant component}
discusses connectivity of giant component; Section \ref{sec:Mobile networks}
discusses recent development, research challenges and opportunities
in mobile networks and Section \ref{sec:Summary} concludes the paper.

\section{Connectivity of Large-Scale Random Networks\label{sec:Large scale network}}

\subsection{Unit disk model and connectivity}

Extensive research has been done on connectivity problems using the
well-known \emph{random geometric graph} and the \emph{unit disk model},
which is usually obtained by randomly and uniformly distributing $n$
nodes in a given area and connecting any two nodes iff their  Euclidean
distance is smaller than or equal to a given threshold $r(n)$, known
as the \emph{transmission range} \cite{Penrose99On,Penrose03Random}.
Significant outcomes have been achieved for both asymptotically infinite
$n$ \cite{Gupta98Critical,Xue04The,Philips89Connectivity,Ravelomanana04Extremal,Balister05Connectivity,Penrose03Random,Balister09A}
and finite $n$ \cite{Bettstetter04On,Bettstetter02On,Tang03An}. 

Research on the connectivity of large-scale random ad-hoc networks
under the unit disk model is spearheaded by Penrose \cite{Penrose97The,Penrose99A}
and Gupta and Kumar \cite{Gupta98Critical}. Specifically, Penrose
\cite{Penrose97The,Penrose99A} and Gupta and Kumar \cite{Gupta98Critical}
proved using different techniques that if the transmission range is
set to $r\left(n\right)=\sqrt{\frac{\log n+c\left(n\right)}{\pi n}}$,
a random network formed by uniformly placing $n$ nodes on a unit-area
disk in $\Re^{2}$ is \emph{asymptotically almost surely} (a.a.s.)
connected as $n\rightarrow\infty$ iff $c\left(n\right)\rightarrow\infty$.
An event $\xi_{n}$ depending on $n$ is said to occur a.a.s. if its
probability tends to one as $n\rightarrow\infty$. Penrose's result
is based on the fact that in the above random network, as $n\rightarrow\infty$
the longest edge of the minimum spanning tree converges in probability
to the minimum transmission range required for the above random network
to have no isolated nodes (or equivalently the longest edge of the
nearest neighbor graph of the above network) \cite{Penrose97The,Penrose99A,Penrose03Random}.
Gupta and Kumar's result is based on a key finding in the continuum
percolation theory \cite[Chapter 6]{Meester96Continuum}: Consider
an infinite network with nodes distributed in $\Re^{2}$ following
a Poisson distribution with density $\rho$; and a pair of nodes separated
by a Euclidean distance $x$ are directly connected with probability
$g\left(x\right)$, independent of the event that another distinct
pair of nodes are directly connected. Here, $g:\Re^{+}\rightarrow\left[0,1\right]$
satisfies the conditions of non-increasing monotonicity and integral
boundedness \cite[pp. 151-152]{Meester96Continuum}. As $\rho\rightarrow\infty$,
a.a.s. the above network in $\Re^{2}$ has only a unique unbounded
component and isolated nodes.

In \cite{Philips89Connectivity}, Philips et al. proved that the average
node degree, i.e. the expected number of neighbors of an arbitrary
node, must grow logarithmically with the area of the network to ensure
that the network is connected, where nodes are placed randomly on
a square according to a Poisson point process with a constant density.
This result by Philips et al. actually provides a necessary condition
on the average node degree required for connectivity. In \cite{Xue04The},
Xue et al. showed that in a network with a total of $n$ nodes randomly
and uniformly distributed on a unit square, if each node is connected
to $c\log n$ nearest neighbors with $c\leq0.074$ then the resulting
random network is a.a.s. disconnected as $n\rightarrow\infty$; and
if each node is connected to $c\log n$ nearest neighbors with $c\geq5.1774$
then the network is a.a.s. connected as $n\rightarrow\infty$. In
\cite{Balister05Connectivity}, Balister et al. advanced the results
in \cite{Xue04The} and improved the lower and upper bounds to $0.3043\log n$
and $0.5139\log n$ respectively. In a more recent paper \cite{Balister09A}
Balister et al. achieved much improved results by showing that there
exists a constant $c_{crit}$ such that if each node is connected
to $\left\lfloor c\log n\right\rfloor $ nearest neighbors with $c<c_{crit}$
then the network is a.a.s. disconnected as $n\rightarrow\infty$,
and if each node is connected to $\left\lfloor c\log n\right\rfloor $
nearest neighbors with $c>c_{crit}$ then the network is a.a.s. connected
as $n\rightarrow\infty$. In both \cite{Balister05Connectivity} and
\cite{Balister09A}, the authors considered nodes randomly distributed
following a Poisson process of intensity one on a square of area $n$.
In \cite{Ravelomanana04Extremal}, Ravelomanana investigated the critical
transmission range for connectivity in 3-dimensional wireless sensor
networks and derived similar results as the 2-dimensional results
in \cite{Gupta98Critical}. 

In \cite{Bettstetter02On}, Bettstetter empirically investigated the
minimum node degree and connectivity of a finite network with $n$
($100\leq n\leq2000$) nodes randomly and uniformly placed on a square
of area $A$. Tang et al. \cite{Tang03An} proposed an empirical formula
relating the probability of having a connected network to the transmission
range for a finite network with $n$ ($n\leq125$) nodes randomly
and uniformly distributed on a unit square. Bettstetter \cite{Bettstetter04On}
studied the network connectivity considering different node placement
models, i.e. uniform distribution, Gaussian distribution. Note that
most results for finite $n$ are empirical results.

\subsection{More general connection models and connectivity}

All the work described in the last subsection is based on the unit
disk model. This model may simplify analysis but no real antenna has
an antenna pattern similar to it. The log-normal shadowing connection
model, which is more realistic than the unit disk model, has accordingly
been considered for investigating network connectivity in \cite{Hekmat06Connectivity,Orriss03Probability,Miorandi05Coverage,Miorandi08The,Bettstetter04failure,Bettstetter05Connectivity}.
Under the log-normal shadowing connection model, two nodes are directly
connected if the received power at one node from the other node, whose
attenuation follows the log-normal model \cite{Rappaport02Wireless},
is greater than or equal to a given threshold.

In \cite{Hekmat06Connectivity}, Hekmat et al. proposed an empirical
formula for computing the average size of the largest connected component
through simulations, where a total of $n$ nodes are randomly and
uniformly distributed in a bounded area in $\Re^{2}$. In \cite{Bettstetter04failure},
Bettstetter derived a lower bound on the minimum node density $\rho$
required to ensure that a network with nodes Poissonly distributed
in an area in $\Re^{2}$ with density $\rho$ is $k$-connected with
a high probability. The analysis is based on the observation that
the minimum node density required for a $k$-connected network is
larger than that required for the network to have a minimum node degree
$k$, and the assumption that the event that a node has a degree greater
than or equal to $k$ is independent of the event that another node
has a degree greater than or equal to $k$. Using simulations, they
showed that the bound is tight when the node density is sufficiently
large. Using the same model as in \cite{Bettstetter04failure}, Bettstetter
et al. obtained in \cite{Bettstetter05Connectivity} a lower bound
on the minimum node density required for an almost surely connected
network using essentially the same technique as that in \cite{Bettstetter04failure}.
The analysis relies on the assumption that the event that a node is
isolated and the event that another node is isolated are independent,
hereafter referred to as the \emph{independence} assumption. Orriss
et al. \cite{Orriss03Probability} considered nodes uniformly and
randomly distributed on a plane and communicating with each other
following the log-normal shadowing model in the framework of cellular
networks. They investigated the distribution of the number of base
stations that communicate with a given mobile and found that the number
of base stations able to communicate with a given mobile \emph{and}
lying within a specified range of the mobile follows a Poisson distribution.
In \cite{Miorandi08The}, Miorandi et al. presented an analytical
procedure for computing the node isolation probability in the presence
of channel randomness, where nodes are distributed following a Poisson
point process in $\Re^{2}$ (which extends their earlier work in \cite{Miorandi05Coverage}).
They further obtained an estimate of the probability that there is
no isolated node in the network based on the above independence assumption.
The previous results in \cite{Hekmat06Connectivity,Orriss03Probability,Miorandi05Coverage,Miorandi08The,Bettstetter04failure,Bettstetter05Connectivity}
dealing with a necessary condition on the critical transmission power
for connectivity under the log-normal shadowing model all rely on
the\emph{ }independence\emph{ }assumption that the node isolation
events are independent. Realistically however, one may expect the
event that a node is isolated and the event that another node is isolated
will be correlated whenever there is a non-zero probability that a
third node may exist which may have direct connections to both nodes.
In the unit disk model, this may happen when the transmission range
of the two nodes overlaps. In the log-normal model, \emph{any} node
may have a non-zero probability of having direct connections to both
nodes. This observation and a lack of rigorous analysis on the node
isolation events to support the independence assumption raised a question
mark over the validity of the results of \cite{Hekmat06Connectivity,Orriss03Probability,Miorandi05Coverage,Miorandi08The,Bettstetter04failure,Bettstetter05Connectivity}. 

Other work in the area includes \cite{Dousse05Impact,Goeckel09Asymptotic,Li09Asymptotic,Kong08Connectivity},
which studies from the percolation perspective, the impact of mutual
interference caused by simultaneous transmissions, the impact of physical
layer cooperative transmissions, the impact of directional antennas
and the impact of unreliable links on connectivity respectively.

\subsection{Random connection model and connectivity}

In the more recent work \cite{Mao11ConnectivityIsolated,Mao11ConnectivityInfinite,Mao10Towards},
the authors considered a network where all nodes are distributed on
a unit square $A\triangleq\left[-\frac{1}{2},\frac{1}{2}\right]^{2}$
following a Poisson distribution with known density $\rho$ and a
pair of nodes are directly connected following a \emph{random connection
model}, viz. a pair of nodes separated by a Euclidean distance $x$
are directly connected with probability $g_{r_{\rho}}\left(x\right)\triangleq g\left(\frac{x}{r_{\rho}}\right)$,
where $g:\left[0,\infty\right)\rightarrow\left[0,1\right]$, independent
of the event that another pair of nodes are directly connected. Here
\begin{equation}
r_{\rho}=\sqrt{\frac{\log\rho+b}{C\rho}}\label{eq:definition of r_rho}
\end{equation}
and $b$ is a constant. The function $g$ is required to satisfy the
properties of non-increasing monotonicity and integral boundedness
\cite[Chapter 6]{Franceschetti07Random,Meester96Continuum}. Further,
it is required that $g$ satisfies the more restrictive requirement
that 
\begin{equation}
g\left(x\right)=o_{x}\left(\frac{1}{x^{2}\log^{2}x}\right)\label{eq:Condition on g(x) requirement 2}
\end{equation}
in order for the impact of the \emph{truncation effect}, which accounts
for the difference between an infinite network and a finite (or asymptotically
infinite) network, on connectivity to be asymptotically vanishingly
small \cite{Mao10Towards}. Based on the above model, it is shown
that as $\rho\rightarrow\infty$, the probability that the above network
has no isolated nodes and the probability that the above network forms
a connected network both converge to $e^{-e^{-b}}$ as $\rho\rightarrow\infty$.
As a ready consequence of these results, the above network is a.a.s.
connected iff $b\rightarrow\infty$ as $\rho\rightarrow\infty$; and
is a.a.s. disconnected iff $b\rightarrow-\infty$ as $\rho\rightarrow\infty$.

The above results extend the earlier work by Penrose \cite{Penrose97The,Penrose99A}
and Gupta and Kumar \cite{Gupta98Critical} from the unit disk model
to the more generic random connection model and bring theoretical
research in the area closer to reality. It can be readily shown that
the results on the random connection model include the work of Penrose
\cite{Penrose97The,Penrose99A} and Gupta and Kumar \cite{Gupta98Critical}
on the unit disk model and the work on the log-normal model \cite{Hekmat06Connectivity,Orriss03Probability,Miorandi05Coverage,Miorandi08The,Bettstetter04failure,Bettstetter05Connectivity}
as two special cases.

\subsection{Challenges}

There remain significant challenges ahead.

Most results in the area rely on three main assumptions: a) the connection
function $g$ is isotropic, b) the connections are independent, c)
nodes are Poissonly or uniformly distributed. 

We conjecture that assumption a) is not a critical assumption, i.e.
under some mild conditions, e.g. nodes are independently and randomly
oriented, assumption a) can be removed while the above results, particularly
the ones obtained assuming a random connection model, are still valid.
It however remains to validate the conjecture. 

The above results however critically rely on assumption b), which
is not necessarily valid in some networks due to channel correlation
and interference, where the latter effect makes the connection between
a pair of nodes dependent on the locations and activities of other
nearby nodes. In \cite{Yang11connectivity}, some preliminary work
was conducted on the connectivity of CSMA networks considering the
impact of interference. The work essentially uses a de-coupling approach
to solve the challenges of connection correlation caused by interference
and suggests that when some realistic constraints are considered,
i.e. carrier-sensing, the connectivity results will be very close
to those obtained under a unit disk model. This conclusion is in stark
contrast with that obtained under an ALOHA multiple-access protocol
\cite{Dousse05Impact}. The major obstacle in dealing with the impact
of channel correlation is that there is no widely accepted model in
the wireless communication community capturing the impact of channel
correlation on connections. 

Finally, it is a logical move after the above work to consider connectivity
of networks with nodes distributed following a generic distribution
other than Poisson or uniform. This remains a major challenge in the
area.

\section{Connectivity of Giant Component\label{sec:Giant component}}

A \emph{giant component} is a component with a designated large percentage
of nodes in the network, say $p$ where $0.5<p<1$. A \emph{component}
is a maximal set of nodes where there is a path between any pair of
nodes in the set. 

Results on connectivity of large-scale random networks under both
the unit disk model \cite{Penrose97The,Penrose99A,Gupta98Critical}
and the more generic random connection model \cite{Mao11ConnectivityIsolated,Mao11ConnectivityInfinite}
revealed the same scaling law. That is, when the number of nodes,
denoted by $n$, in a network increases, the transmission range (or
power) has to increase at a rate to maintain an average node degree
of $\Theta\left(\log n\right)$ in order to achieve connectivity.
For two functions $f\left(x\right)$ and $h\left(x\right)$, $f\left(x\right)=\Theta\left(h\left(x\right)\right)$
iff there exist a sufficiently large $x_{0}$ and two positive constants
$c_{1}$ and $c_{2}$ such that for any $x>x_{0}$, $c_{1}h\left(x\right)\geq f\left(x\right)\geq c_{2}h\left(x\right)$.
For example, the critical transmission range for connectivity is $r\left(n\right)=\sqrt{\frac{\log n+c\left(n\right)}{\pi n}}$
under the unit disk model for a random network formed by uniformly
placing $n$ nodes on a unit-area disk \cite{Penrose97The,Penrose99A,Gupta98Critical}%
\footnote{By scaling, it can be shown that assuming an extended network model
where nodes are distributed on a disk of area $n$ with a constant
density of $1$ node per unit area, the critical transmission range
for connectivity is $r\left(n\right)=\sqrt{\frac{\log n+c\left(n\right)}{\pi}}$.%
}. In other words, a connected network poses a very demanding requirement
on the transmission range (or power). This in turn causes many undesirable
effects on increased interference and reduced throughput. In \cite{Gupta00the},
it was shown that the end-to-end throughput between a randomly chosen
source-destination pair in the above network is $\Theta\left(\frac{W}{\sqrt{n\log n}}\right)$,
where $W$ is the link capacity. This result can be intuitively explained
using the results on connectivity as follows: as the number of nodes
$n$ increases, the average distance, measured by the number of hops,
between a randomly chosen pair of nodes is $\Theta\left(\frac{1}{r\left(n\right)}\right)=\Theta\left(\sqrt{\frac{\pi n}{\log n}}\right)$.
That is, for a \emph{typical} node, for every packet transmitted for
itself, there are $\Theta\left(\sqrt{\frac{\pi n}{\log n}}\right)$
relay packets transmitted for other source-destination pairs. Further,
the average node degree is $n\pi r^{2}\left(n\right)=\Theta\left(\log n\right)$,
which implies that in a neighborhood of a typical node, at any time
there can only be one out of every $\Theta\left(\log n\right)$ nodes
active. It follows that the end-to-end throughput between a typical
source-destination pair is $\frac{W}{\Theta\left(\sqrt{\frac{\pi n}{\log n}}\right)\Theta\left(\log n\right)}=\Theta\left(\frac{W}{\sqrt{n\log n}}\right)$,
hence comes the result in \cite{Gupta00the}.

The above observation motivates a question: since the network connectivity
is a very demanding requirement, whether there is any benefit in backing
down from such a demanding requirement and requiring most nodes, instead
all nodes, to be connected?

Indeed in many applications, it is unnecessary for all nodes to always
be connected to each other \cite{Blough06The}. Examples of such applications
include a wireless sensor network for habitat monitoring \cite{Hu05The,Juang02Energy}
or environmental monitoring \cite{Padhy05Glacial,Ingraham05Wireless}
and a mobile ad-hoc network in which users can tolerate short off-service
intervals \cite{Yu08Delay}.

In environmental monitoring, there are scenarios where the size of
the monitored phenomenon is very large (e.g. rain clouds) or the parameters
(e.g. temperature, humidity) that are monitored change slowly both
in space and in time. When the number of nodes for monitoring the
phenomenon or measuring the parameters is very large, having a few
disconnected nodes will not cause a statistically significant change
in the monitored parameters. One example of such applications is a
wireless sensor network that was deployed underneath the Briksdalsbreen
glacier in Norway to monitor the pressure, humidity, and temperature
of ice to understand glacial dynamics in response to climate change
\cite{Padhy05Glacial}. In habitat monitoring, there are scenarios
where the number of objects (e.g. zebras and cane toads \cite{Hu05The})
that are monitored is large. Having a few nodes disconnected or lost
may not significantly affect the accuracy of the monitored parameter.
In many mobile ad-hoc networks, having a number of nodes temporarily
disconnected is also not critical, as long as users can tolerate short
off-service intervals. For example, in a campus-wide wireless network,
students and staff can share information using wireless devices (e.g.
laptops and personal digital assistants) around the campus \cite{Yu08Delay}.
When a wireless device temporarily loses connection, it can store
the data and complete the work after becoming connected later.

In \cite{Ta09On,Ta09OnDisk}, considering a network with a total of
$n$ nodes uniformly and i.i.d. on a unit square in $\Re^{2}$, it
was shown analytically that under both the unit disk model \cite{Ta09OnDisk}
and the log-normal model \cite{Ta09On}, the transmission range (or
power) required for having a designated large percentage of nodes
connected, say $p$ where $0.5\leq p<1$, is asymptotically vanishingly
small compared to that required for having a connected network, irrespective
of the value of $p$. This result implies that significant energy
savings can be achieved if we require only most nodes (e.g. $95\%$,
$99\%$) to be connected, instead of requiring all nodes to be connected;
and given a network with most nodes connected, a sharp increase in
the transmission range (or power) is required to connect the few remaining
hard-to-reach nodes. It was further shown using simulations that under
the unit disk model, in a network with $1000$ nodes, the transmission
range required for having $95\%$ nodes connected is only $76\%$
of that required for having all nodes connected. Based on a conservative
estimate that the required transmission power increases with the square
of the required transmission range, an energy saving of at least $42\%$
can be achieved by sacrificing $5\%$ of nodes. That energy saving
will further increase with an increase in the number of nodes in the
network. Other benefits of the reduced transmission range or power
requirement is the reduced interference, hence better throughput.

It remains to find the value of the transmission range (or power)
required for guaranteeing a designated large percentage of nodes to
be connected in a large scale network. This problem has some intrinsic
connections to the problem of finding the percolation probability
in the continuum percolation theory \cite{Meester96Continuum}. Further,
it remains to quantitatively characterize the benefit in capacity
due to the reduced transmission range (or power) required for a giant
component.

Other researchers approached the problem caused by the demanding requirement
of a connected network on the transmission range (or power) from a
different perspective and considered the use of infrastructure instead.
Here the infrastructure can be quite flexible. It can be a subset
of nodes connected through wired connections \cite{Liu03On}, or a
subset of nodes with possibly more powerful transmission capability
that forms a wireless backbone of the network \cite{Franceschetti07Closing,Li10The},
or a subset of nodes with satellite links as one would possibly encounter
in UAV formations\cite{Mao07Design}. The use of infrastructure does
not change the wireless multi-hop nature of the end-to-end communication,
instead the infrastructure assists the end-to-end communication by
leapfrogging some long hops and reducing the number of hops between
two nodes, hence improving the performance. Accordingly the concept
of k-hop connected networks was proposed and investigated \cite{Wang09Mobility,Mao10Probability,Ng11Analysis,Ta07Evaluation}.
In a k-hop connected network, the maximum number of hops between any
two nodes is smaller than or equal to $k$. Some research in the area
was also conducted under the name of hybrid networks \cite{Dousse02Connectivity,Liu03On}.

Despite previous research in the area of hybrid networks or k-hop
connected networks, no conclusive results have been obtained yet on
the role of infrastructure in wireless multi-hop networks with many
problems remain unanswered. Some examples include: for randomly deployed
infrastructure nodes and {}``ordinary'' nodes, how many infrastructure
nodes (versus ordinary nodes) are required for a k-hop connected network;
for deterministically deployed infrastructure nodes and randomly deployed
ordinary nodes, how many infrastructure nodes are required for a k-hop
connected network and what is the optimum deployment of infrastructure
nodes; how to combine the use of infrastructure-based communications
and ad-hoc communications in one network in order to provide some
performance guarantee, in terms of capacity or delay. These problems
are important for wireless multi-hop networks, particularly for wireless
vehicular networks in which both infrastructure-based communications
and ad-hoc communications will co-exist \cite{Yang09Relay}.

\section{Development and Challenges in Mobile Networks\label{sec:Mobile networks}}

In \cite{Grossglauser02Mobility}, Grossglauser and Tse studied the
capacity of mobile ad-hoc networks. Particularly, they considered
a network with a total of $n$ nodes distributed on a unit-area disk,
the trajectories of different nodes are i.i.d. and the nodes' movement
is such that the spatial distribution of nodes are stationary and
ergodic with stationary uniform distribution on the disk. They showed
that in the above network with \emph{unbounded delay requirement},
the throughput between a randomly chosen source-destination pair can
be kept \emph{constant} even as $n$ increases. This result is in
stark contrast with its counter-part in static networks in which the
throughout between a randomly chosen source-destination pair is shown
to be $\Theta\left(\frac{W}{\sqrt{n\log n}}\right)$ \cite{Gupta00the}.
Following the seminal work of Grossglauser and Tse, other researchers
have conducted further research trying to quantitatively characterize
the relationship between delay, mobility and capacity in mobile ad-hoc
networks \cite{Sharma07Delay,Neely05Capacity,Wang09Mobility,Gamal04Throughput,Toumpis04Large}
and the obtained results vary greatly with the mobility models and
network settings.

A fundamental reason why mobility increases throughput is that in
mobile networks message transmissions generally follow the store-carry-forward
pattern versus the store-forward pattern found in static networks.
As nodes move, new opportunity may arise such that a mobile node can
carry the message until it meets a node, which is in a better position
than itself to transmit the message to the destination, or until it
meets the destination directly. In this way, the number of relay nodes
(number of hops) involved in transmitting a message to its destination
can be greatly reduced and the required transmission range (or power)
for a node to reach another node via a multi-hop path can also be
greatly reduced, hence the benefit in improved capacity. The cost
in achieving this benefit in capacity is the increased delay.

By analogy, mobility can also improve connectivity. There are three
fundamental differences between mobile networks and static networks
\cite{Mao09Graph}: in mobile networks
\begin{itemize}
\item the wireless link between two directly connected nodes and the end-to-end
path only exists temporarily; 
\item two nodes may never be part of the same connected component but they
are still able to communicate, i.e. exchange messages, with each other;
and 
\item while any one wireless link may be (or assumed to be) undirectional,
the path connecting any two nodes is directional, i.e. there is a
path from node $v_{i}$ to node $v_{j}$ within a designated time
period does not necessarily mean there is a path from $v_{j}$ to
$v_{i}$ within the same period.
\end{itemize}
\begin{figure*}
\begin{centering}
\includegraphics[width=0.6\textwidth]{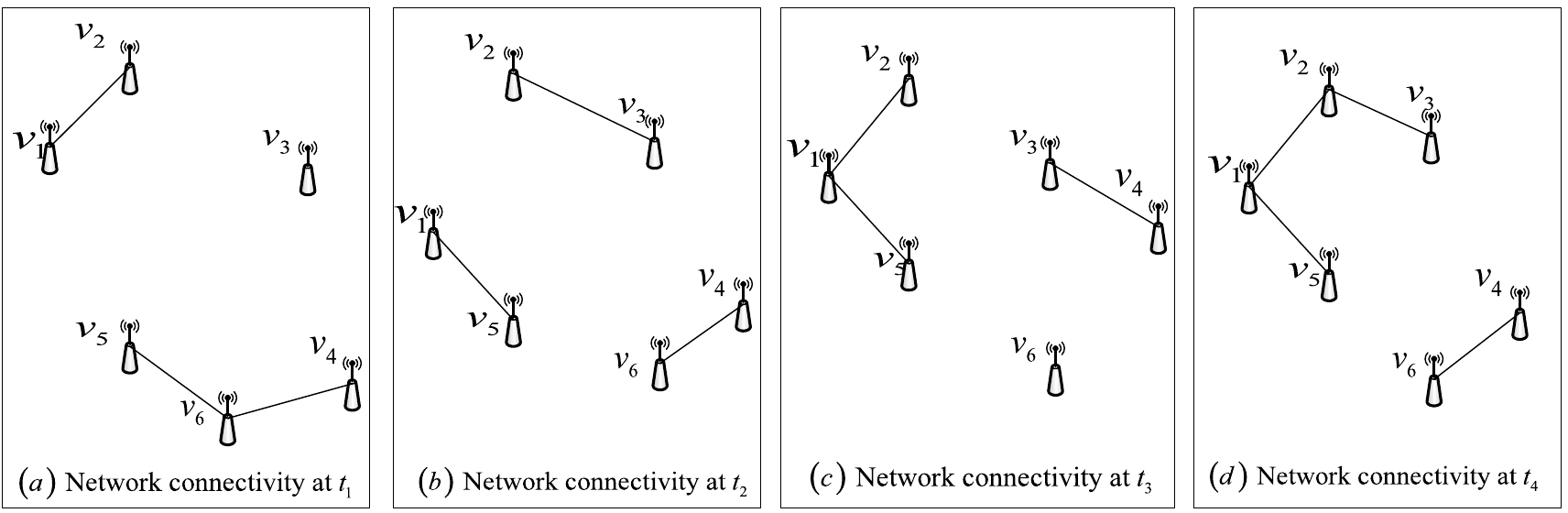}
\par\end{centering}

\caption{An illustration of connectivity in a mobile ad-hoc network. A solid
line represents a connection between two nodes. The network is disconnected
at any time instant but there is a path from any node to any other
node in the network. For example, nodes $v_{1}$ and $v_{6}$ are
never part of the same connected component but a message from $v_{1}$
can still reach $v_{6}$ through the following path: $t_{1}:v_{1}\rightarrow v_{2}$,
$t_{2}:v_{2}\rightarrow v_{3}$, $t_{3}:v_{3}\rightarrow v_{4}$,
$t_{4}:v_{4}\rightarrow v_{6}$. Further, a message from $v_{6}$
can reach $v_{1}$ at $t_{2}$ but a message from $v_{1}$ can only
reach $v_{6}$ at $t_{4}$.\label{fig:mobile ad hoc network connectivity}}
\end{figure*}

These are illustrated in Fig. \ref{fig:mobile ad hoc network connectivity}.
Particularly the last difference implies that it is important to consider
the order of links in time when analyzing mobile networks, which has
been incorrectly neglected in some previous work.

Due to these differences, many established concepts in static networks
must be revisited for mobile networks. For example, a static wireless
multi-hop network is said to be connected iff there is a path between
any pair of nodes in the network. However a more meaningful definition
of connectivity in mobile networks is to say that a mobile network
is connected in time period $[0,T]$ if any node can exchange a message
with any other node within $[0,T]$. The above definition implies
that the tradeoff between connectivity, mobility and delay is the
prime issue when analyzing the connectivity of mobile networks. Despite
intensive research on the properties of mobile networks, no conclusive
results have been obtained on the above problem and it remains a major
challenge in the area.

\section{Summary \label{sec:Summary}}

Wireless multi-hop networks have attracted significant research interest.
This interest is expected to grow further with the proliferation of
applications, particularly in the areas of wireless vehicular networks
and sensor networks. In this paper, we briefly overviewed recent development
and discussed research challenges and opportunities in the area mainly
from the perspective of network connectivity. We also showed how the
results on network connectivity is related the study of other performance
metrics, i.e. capacity and delay.

\bibliographystyle{IEEEtran}

\begin{thebibliography}{10}
\providecommand{\url}[1]{#1}
\csname url@samestyle\endcsname
\providecommand{\newblock}{\relax}
\providecommand{\bibinfo}[2]{#2}
\providecommand{\BIBentrySTDinterwordspacing}{\spaceskip=0pt\relax}
\providecommand{\BIBentryALTinterwordstretchfactor}{4}
\providecommand{\BIBentryALTinterwordspacing}{\spaceskip=\fontdimen2\font plus
\BIBentryALTinterwordstretchfactor\fontdimen3\font minus
  \fontdimen4\font\relax}
\providecommand{\BIBforeignlanguage}[2]{{%
\expandafter\ifx\csname l@#1\endcsname\relax
\typeout{** WARNING: IEEEtran.bst: No hyphenation pattern has been}%
\typeout{** loaded for the language `#1'. Using the pattern for}%
\typeout{** the default language instead.}%
\else
\language=\csname l@#1\endcsname
\fi
#2}}
\providecommand{\BIBdecl}{\relax}
\BIBdecl

\bibitem{Gupta98Critical}
P.~Gupta and P.~R. Kumar, \emph{Critical Power for Asymptotic Connectivity in
  Wireless Networks}.\hskip 1em plus 0.5em minus 0.4em\relax Boston, MA:
  Birkhauser, 1998, pp. 547--566.

\bibitem{Haenggi09Stochastic}
M.~Haenggi, J.~G. Andrews, F.~Baccelli, O.~Dousse, and M.~Franceschetti,
  ``Stochastic geometry and random graphs for the analysis and design of
  wireless networks,'' \emph{IEEE Journal on Selected Areas in Communications},
  vol.~27, no.~7, pp. 1029--1046, 2009.

\bibitem{Penrose03Random}
M.~D. Penrose, \emph{Random Geometric Graphs}, ser. Oxford Studies in
  Probability.\hskip 1em plus 0.5em minus 0.4em\relax Oxford University Press,
  USA, 2003.

\bibitem{Penrose99On}
------, ``On k-connectivity for a geometric random graph,'' \emph{Random
  Structures and Algorithms}, vol.~15, no.~2, pp. 145--164, 1999.

\bibitem{Xue04The}
F.~Xue and P.~Kumar, ``The number of neighbors needed for connectivity of
  wireless networks,'' \emph{Wireless Networks}, vol.~10, no.~2, pp. 169--181,
  2004.

\bibitem{Philips89Connectivity}
T.~K. Philips, S.~S. Panwar, and A.~N. Tantawi, ``Connectivity properties of a
  packet radio network model,'' \emph{IEEE Transactions on Information Theory},
  vol.~35, no.~5, pp. 1044--1047, 1989.

\bibitem{Ravelomanana04Extremal}
V.~Ravelomanana, ``Extremal properties of three-dimensional sensor networks
  with applications,'' \emph{IEEE Transactions on Mobile Computing}, vol.~3,
  no.~3, pp. 246--257, 2004.

\bibitem{Balister05Connectivity}
P.~Balister, B.~Bollobas, A.~Sarkar, and M.~Walters, ``Connectivity of random
  k-nearest-neighbour graphs,'' \emph{Advances in Applied Probability},
  vol.~37, no.~1, pp. 1--24, 2005.

\bibitem{Balister09A}
------, ``A critical constant for the k nearest neighbour model,''
  \emph{Advances in Applied Probability}, vol.~41, no.~1, pp. 1--12, 2009.

\bibitem{Bettstetter04On}
C.~Bettstetter, ``On the connectivity of ad hoc networks,'' \emph{The Computer
  Journal}, vol.~47, no.~4, pp. 432--447, 2004.

\bibitem{Bettstetter02On}
------, ``On the minimum node degree and connectivity of a wireless multihop
  network,'' in \emph{3rd ACM International Symposium on Mobile Ad Hoc
  Networking and Computing}, pp. 80--91.

\bibitem{Tang03An}
A.~Tang, C.~Florens, and S.~H. Low, ``An empirical study on the connectivity of
  ad hoc networks,'' in \emph{IEEE Aerospace Conference}, vol.~3, pp.
  1333--1338.

\bibitem{Penrose97The}
M.~Penrose, ``The longest edge of the random minimal spanning tree,'' \emph{The
  Annals of Applied Probability}, vol.~7, no.~2, pp. 340--361, 1997.

\bibitem{Penrose99A}
------, ``A strong law for the longest edge of the minimal spanning tree,''
  \emph{The Annals of Applied Probability}, vol.~27, no.~1, pp. 246--260, 1999.

\bibitem{Meester96Continuum}
R.~Meester and R.~Roy, \emph{Continuum Percolation}.\hskip 1em plus 0.5em minus
  0.4em\relax Cambridge University Press, 1996.

\bibitem{Hekmat06Connectivity}
R.~Hekmat and P.~V. Mieghem, ``Connectivity in wireless ad-hoc networks with a
  log-normal radio model,'' \emph{Mobile Networks and Applications}, vol.~11,
  no.~3, pp. 351--360, 2006.

\bibitem{Orriss03Probability}
J.~Orriss and S.~K. Barton, ``Probability distributions for the number of radio
  transceivers which can communicate with one another,'' \emph{IEEE
  Transactions on Communications}, vol.~51, no.~4, pp. 676--681, 2003.

\bibitem{Miorandi05Coverage}
D.~Miorandi and E.~Altman, ``Coverage and connectivity of ad hoc networks
  presence of channel randomness,'' in \emph{IEEE INFOCOM}, vol.~1, pp.
  491--502.

\bibitem{Miorandi08The}
D.~Miorandi, ``The impact of channel randomness on coverage and connectivity of
  ad hoc and sensor networks,'' \emph{IEEE Transactions on Wireless
  Communications}, vol.~7, no.~3, pp. 1062--1072, 2008.

\bibitem{Bettstetter04failure}
C.~Bettstetter, ``Failure-resilient ad hoc and sensor networks in a shadow
  fading environment,'' in \emph{IEEE/IFIP International Conference on
  Dependable Systems and Networks}.

\bibitem{Bettstetter05Connectivity}
C.~Bettstetter and C.~Hartmann, ``Connectivity of wireless multihop networks in
  a shadow fading environment,'' \emph{Wireless Networks}, vol.~11, no.~5, pp.
  571--579, 2005.

\bibitem{Rappaport02Wireless}
T.~S. Rappaport, \emph{Wireless Communications: Principles and Practice}.\hskip
  1em plus 0.5em minus 0.4em\relax Prentice Hall, 2002.

\bibitem{Dousse05Impact}
O.~Dousse, F.~Baccelli, and P.~Thiran, ``Impact of interferences on
  connectivity in ad hoc networks,'' \emph{IEEE/ACM Transactions on
  Networking}, vol.~13, no.~2, pp. 425--436, 2005.

\bibitem{Goeckel09Asymptotic}
D.~Goeckel, L.~Benyuan, D.~Towsley, W.~Liaoruo, and C.~Westphal, ``Asymptotic
  connectivity properties of cooperative wireless ad hoc networks,'' \emph{IEEE
  Journal on Selected Areas in Communications}, vol.~27, no.~7, pp. 1226--1237,
  2009.

\bibitem{Li09Asymptotic}
P.~Li, C.~Zhang, and Y.~Fang, ``Asymptotic connectivity in wireless ad hoc
  networks using directional antennas,'' \emph{IEEE/ACM Transactions on
  Networking}, vol.~17, no.~4, pp. 1106--1117, 2009.

\bibitem{Kong08Connectivity}
Z.~Kong and E.~M. Yeh, ``Connectivity and latency in large-scale wireless
  networks with unreliable links,'' in \emph{IEEE INFOCOM}, pp. 11--15.

\bibitem{Mao11ConnectivityIsolated}
G.~Mao and B.~D. Anderson, ``Connectivity of large scale networks: Distribution
  of isolated nodes,'' \emph{submitted to IEEE Transactions on Mobile
  Computing, available at http://arxiv.org/abs/1103.1994}, 2011.

\bibitem{Mao11ConnectivityInfinite}
------, ``Connectivity of large scale networks: Emergence of unique unbounded
  component,'' \emph{submitted to IEEE Transactions on Mobile Computing,
  available at http://arxiv.org/abs/1103.1991}, 2011.

\bibitem{Mao10Towards}
------, ``Towards a better understanding of large scale network models,''
  \emph{accepted to appear in IEEE/ACM Transactions on Networking}, 2011.

\bibitem{Franceschetti07Random}
M.~Franceschetti and R.~Meester, \emph{Random Networks for
  Communication}.\hskip 1em plus 0.5em minus 0.4em\relax Cambridge University
  Press, 2007.

\bibitem{Yang11connectivity}
T.~Yang, G.~Mao, and W.~Zhang, ``Connectivity of wireless csma multi-hop
  networks,'' in \emph{IEEE International Conference on Communications}, pp.
  1--5.

\bibitem{Gupta00the}
P.~Gupta and P.~Kumar, ``The capacity of wireless networks,'' \emph{IEEE
  Transactions on Information Theory}, vol.~46, no.~2, pp. 388--404, 2000.

\bibitem{Blough06The}
D.~M. Blough, M.~Leoncini, G.~Resta, and P.~Santi, ``The k-neighbors approach
  to interference bounded and symmetric topology control in ad hoc networks,''
  \emph{IEEE Transactions on Mobile Computing}, vol.~5, no.~9, pp. 1267 --
  1282, 2006.

\bibitem{Hu05The}
W.~Hu, V.~N. Tran, N.~Bulusu, C.~Chou, S.~Jha, and A.~Taylor, ``The design and
  evaluation of a hybrid sensor network for cane-toad monitoring,'' in
  \emph{Proc. 4th Int. Symp. Inf. Process. Sens. Netw.}, pp. 503 -- 508.

\bibitem{Juang02Energy}
P.~Juang, H.~Oki, Y.~Wang, M.~Martonosi, L.~Peh, and D.~Rubenstein,
  ``Energy-efficient computing for wildlife tracking: Design tradeoffs and
  early experiences with zebranet,'' \emph{ACM SIGARCH Comput. Archit. News},
  vol.~30, no.~5, pp. 96 -- 107, 2002.

\bibitem{Padhy05Glacial}
P.~Padhy, K.~Martinez, A.~Riddoch, J.~K. Hart, and H.~L.~R. Ong, ``Glacial
  environment monitoring using sensor networks,'' in \emph{Proc. 1st REALWSN},
  pp. 10 -- 14.

\bibitem{Ingraham05Wireless}
D.~Ingraham, R.~Beresford, K.~Kaluri, M.~Ndoh, and K.~Srinivasan, ``Wireless
  sensors: Oyster habitat monitoring in the bras d'or lakes,'' in \emph{IEEE
  1st International Conference on Distributed Computing In Sensor Systems}, pp.
  399 -- 400.

\bibitem{Yu08Delay}
X.~Yu and S.~Chandra, ``Delay-tolerant collaborations among campus wide
  wireless users,'' in \emph{IEEE INFOCOM}, pp. 2101 -- 2109.

\bibitem{Ta09On}
X.~Ta, G.~Mao, and B.~D. Anderson, ``On the giant component of wireless
  multi-hop networks in the presence of shadowing,'' \emph{IEEE Transactions on
  Vehicular Technology}, vol.~58, no.~9, pp. 5152--5163, 2009.

\bibitem{Ta09OnDisk}
------, ``On the properties of giant component in wireless multi-hop
  networks,'' in \emph{IEEE INFOCOM}, pp. 2556 -- 2560.

\bibitem{Liu03On}
B.~Liu, Z.~Liu, and D.~Towsley, ``On the capacity of hybrid wireless
  networks,'' in \emph{IEEE INFOCOM}, vol.~2, pp. 1543--1552.

\bibitem{Franceschetti07Closing}
M.~Franceschetti, O.~Dousse, D.~N.~C. Tse, and P.~Thiran, ``Closing the gap in
  the capacity of wireless networks via percolation theory,'' \emph{IEEE
  Transactions on Information Theory}, vol.~53, no.~3, pp. 1009--1018, 2007.

\bibitem{Li10The}
P.~Li and Y.~Fang, ``The capacity of heterogeneous wireless networks,'' in
  \emph{IEEE INFOCOM}, 2010, pp. 1--9.

\bibitem{Mao07Design}
G.~Mao, S.~Drake, and B.~D.~O. Anderson, ``Design of an extended kalman filter
  for uav localization,'' in \emph{Information, Decision and Control}, 2007,
  pp. 224--229.

\bibitem{Wang09Mobility}
Q.~Wang, X.~Wang, and X.~Lin, ``Mobility increases the connectivity of k-hop
  clustered wireless networks,'' in \emph{MobiCom}, 2009, pp. 121 -- 132.

\bibitem{Mao10Probability}
G.~Mao, Z.~Zhang, and B.~Anderson, ``Probability of k-hop connection under
  random connection model,'' \emph{IEEE Communication Letters}, vol.~14,
  no.~11, pp. 1023 -- 1025, 2010.

\bibitem{Ng11Analysis}
S.~C. Ng, W.~Zhang, Y.~Yang, and G.~Mao, ``Analysis of access and connectivity
  probabilities in vehicular relay networks,'' \emph{IEEE Journal on Selected
  Areas in Communications--Special Issue Vehicular Communications and
  Networks}, vol.~29, no.~1, pp. 140 -- 150, 2011.

\bibitem{Ta07Evaluation}
X.~Ta, G.~Mao, and B.~D.~O. Anderson, ``Evaluation of the probability of k-hop
  connection in homogeneous wireless sensor networks,'' in \emph{IEEE
  Globecom}, 2007, pp. 1279 -- 1284.

\bibitem{Dousse02Connectivity}
O.~Dousse, P.~Thiran, and M.~Hasler, ``Connectivity in ad-hoc and hybrid
  networks,'' in \emph{IEEE INFOCOM}, vol.~2, pp. 1079--1088.

\bibitem{Yang09Relay}
Y.~Yang, H.~Hu, J.~Xu, and G.~Mao, ``Relay technologies for wimax and
  lte-advanced mobile systems,'' \emph{IEEE Communications Magazine}, vol.~47,
  no.~10, pp. 100--105, 2009.

\bibitem{Grossglauser02Mobility}
M.~Grossglauser and D.~N.~C. Tse, ``Mobility increases the capacity of ad hoc
  wireless networks,'' \emph{IEEE/ACM Transactions on Networking}, vol.~10,
  no.~4, pp. 477--486, 2002.

\bibitem{Sharma07Delay}
G.~Sharma, R.~Mazumdar, and B.~Shroff, ``Delay and capacity trade-offs in
  mobile ad hoc networks: A global perspective,'' \emph{IEEE/ACM Transactions
  on Networking}, vol.~15, no.~5, pp. 981--992, 2007.

\bibitem{Neely05Capacity}
M.~J. Neely and E.~Modiano, ``Capacity and delay tradeoffs for ad hoc mobile
  networks,'' \emph{IEEE Transactions on Information Theory}, vol.~51, no.~6,
  pp. 1917--1937, 2005.

\bibitem{Gamal04Throughput}
A.~E. Gamal, J.~Mammen, B.~Prabhakar, and D.~Shah, ``Throughput-delay trade-off
  in wireless networks,'' in \emph{IEEE INFOCOM}, vol.~1, 2004.

\bibitem{Toumpis04Large}
S.~Toumpis and A.~J. Goldsmith, ``Large wireless networks under fading,
  mobility, and delay constraints,'' in \emph{IEEE INFOCOM}, vol.~1, 2004, pp.
  609--619.

\bibitem{Mao09Graph}
G.~Mao and B.~D. Anderson, ``Graph theoretic models and tools for the analysis
  of dynamic wireless multihop networks,'' in \emph{IEEE WCNC}, 2009, pp. 1--6.

\end{thebibliography}

\end{document}